\documentclass[aip,jap,reprint,groupeaddress,amssymb]{revtex4-1}
\pdfoutput=1
\usepackage{amsmath}    % need for subequations
\usepackage{graphicx}   % need for figures
\usepackage{verbatim}   % useful for program listings
\usepackage{color}      % use if color is used in text
\usepackage{subfigure}  % use for side-by-side figures
\usepackage{hyperref}   % use for hypertext links, including those to external documents and URLs
\usepackage{soul}

\graphicspath{{./figures/}}

\begin{document}

\title{Top-down fabrication of plasmonic nanostructures for deterministic coupling to single quantum emitters}
\author{Wolfgang Pfaff}
\email{w.pfaff@tudelft.nl}
\author{Arthur Vos}
\author{Ronald Hanson}
\affiliation{Kavli Institute of Nanoscience Delft, Delft University of Technology, P.O. Box 5046, 2600 GA Delft, The Netherlands}

\begin{abstract}
Metal nanostructures can be used to harvest and guide the emission of single photon emitters on-chip via surface plasmon polaritons. In order to develop and characterize photonic devices based on emitter-plasmon hybrid structures a deterministic and scalable fabrication method for such structures is desirable. Here we demonstrate deterministic and scalable top-down fabrication of metal wires onto preselected nitrogen vacancy centers in nanodiamonds using clean room nano-fabrication methods. We observe a life-time reduction of the emitter emission that is consistent with earlier proof-of-principle experiments that used non-deterministic fabrication methods. This result indicates that top-down fabrication is a promising technique for processing future devices featuring single photon emitters and plasmonic nanostructures. 
\end{abstract}

\maketitle

\section{Introduction}

Metal nanostructures can be used to tailor the emission of single photon emitters. For instance, antennas can control emitted radiation into free space \cite{Novotny2011,Curto2010}. Moreover, a substantial fraction of the emission of a single emitter can be captured and guided on-chip by metal wires in the form of surface plasmon polariton (SPP) waveguides \cite{Chang2006,2009NatPh...5..470K,2007Natur.450..402A}. Single SPPs can be detected on-chip using integrated detectors \cite{Heeres2010,Falk2009}, making them a promising tool for integrating single photon emitters into photonic circuits. Most previous fabrication approaches for devices suitable for such integration use either random \cite{2009NatPh...5..470K,2007Natur.450..402A,Falk2009} or one-by-one assembly \cite{Huck2011,Schell:1335271} and are therefore not easily scalable. Scalable device fabrication can be done using techniques based on chemical functionalization\cite{Curto2010} and lithography \cite{2012PSSBR.249..678P,Gruber:2012gi}.

The nitrogen-vacancy (NV) center in diamond is a stable single-photon source at room temperature that has gained much recognition as promising platform for future solid-state quantum technologies \cite{Awschalom2007,Aharonovich:2011cu,Jelezko:2006jq}. Furthermore, diamond nano-crystals containing single NV centers can be integrated into on-chip photonic structures \cite{Benson:2011hx,2010NanoL..10.3922E,Wolters:2010bd,2011ApPhL..98s3103V}. Here we demonstrate an important step towards scalable fabrication of NV center-metal hybrid structures. We deterministically place metal wires onto pre-characterized single emitters in diamond nano-crystals using electron beam lithography (EBL). We observe reduced optical life times of the NV centers after the wire placement, consistent with coupling of the emission to plasmonic modes. Our approach is suited for mass-fabrication of devices, allowing for the development and systematic characterization of photonic circuits that use plasmonic structures to harvest radiation of single photon emitters.

\section{Concept and samples}

\begin{figure}
\includegraphics{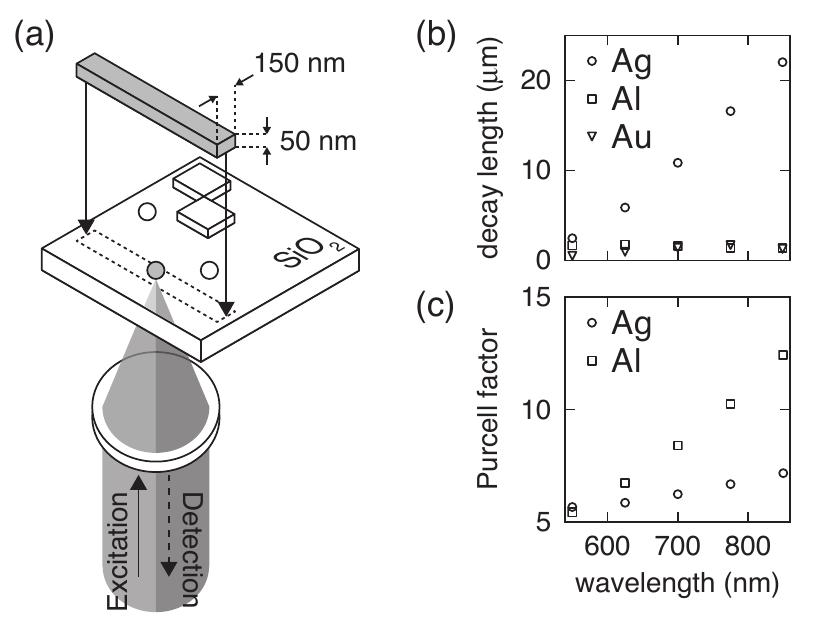}
\caption{\label{fig1}Sample design and measurement approach. {\bf a,} We fabricate a metal nanowire deterministically on top of an NV center contained in a diamond nanocrystal on a glass substrate. Precise localization of the emitter can be done with respect to prefabricated position markers (indicated in the top right of the sample). The emitter can be characterized before and after fabrication in a confocal microscope through the back side of the substrate. {\bf b,} Simulated propagation length of wires from different metals. The simulation does not take into account the graininess of the material resulting from deposition by evaporation. {\bf c,} Simulated emission enhancement of an emitter located underneath a metal wire. For simplicity we ignore the effect of the diamond nano-particle in the simulation. We assume that the wire is placed directly on the glass surface, with the emitter embedded in the substrate $20\,\mathrm{nm}$ underneath the surface and centered under the wire, with a single radiating dipole oriented in vertical direction.}
\end{figure}

The concept of the experiment is illustrated in Fig.\ \ref{fig1}a. We fabricate metal wires with a width of $150\,\mathrm{nm}$, a thickness of $50\,\mathrm{nm}$, and variable length ($2-10\,\mu\mathrm{m}$) onto pre-selected NV centers contained in diamond nanocrystals dispersed on a glass substrate. This allows characterization of emitters before and after wire fabrication, using confocal microscopy through the substrate. The chosen dimensions in principle allow the wire to serve as a single-mode SPP waveguide at the emission length of the NV center \cite{Chang2006,Kusar:2012gp}, making it an appropriate testbed structure. We note, however, that our approach is entirely independent of the precise geometry of the structure fabricated and is only limited by the spatial accuracy with which the emitter can be located and the precision of the EBL patterning.

Several metals support propagation of SPPs at optical frequencies. Figure \ref{fig1}b shows expected propagation lengths for wires of gold, silver and aluminum, with the sample geometry as given above. The results have been obtained using finite difference time domain (FDTD) simulations (Lumerical). These show that silver can be expected to give the best propagation. SPPs are propagated little by aluminum, whereas gold experiences a cut-off of the propagation for wavelengths below $\sim 800\,\mathrm{nm}$
% , commonly attributed to interband transitions causing absorption (REF)
and is therefore not well-suited for waveguides at the emission frequencies of NV centers. We limit our investigations here to wires of silver and aluminum, noting that adaptation for other materials is straight-forward. For the simulated structure, $\sim 60\%$ of the total emission is emitted into guided SPP modes in the case of a silver wire, and $\sim 20\%$ for aluminum. Figure \ref{fig1}c shows estimated emission enhancements for an emitter close to the wire, where we define the Purcell factor as\cite{2009NatPh...5..470K}
\begin{equation}
	F_\mathrm P = \frac{\Gamma}{\Gamma_0},
	\label{eq:purcell}
\end{equation}
where $\Gamma_0$ and $\Gamma$ are the total emission rates before and after the wire placement, respectively. The results are obtained from FDTD simulations using a simplified sample geometry (see caption of Fig.\,\ref{fig1}).

\begin{figure}
\includegraphics{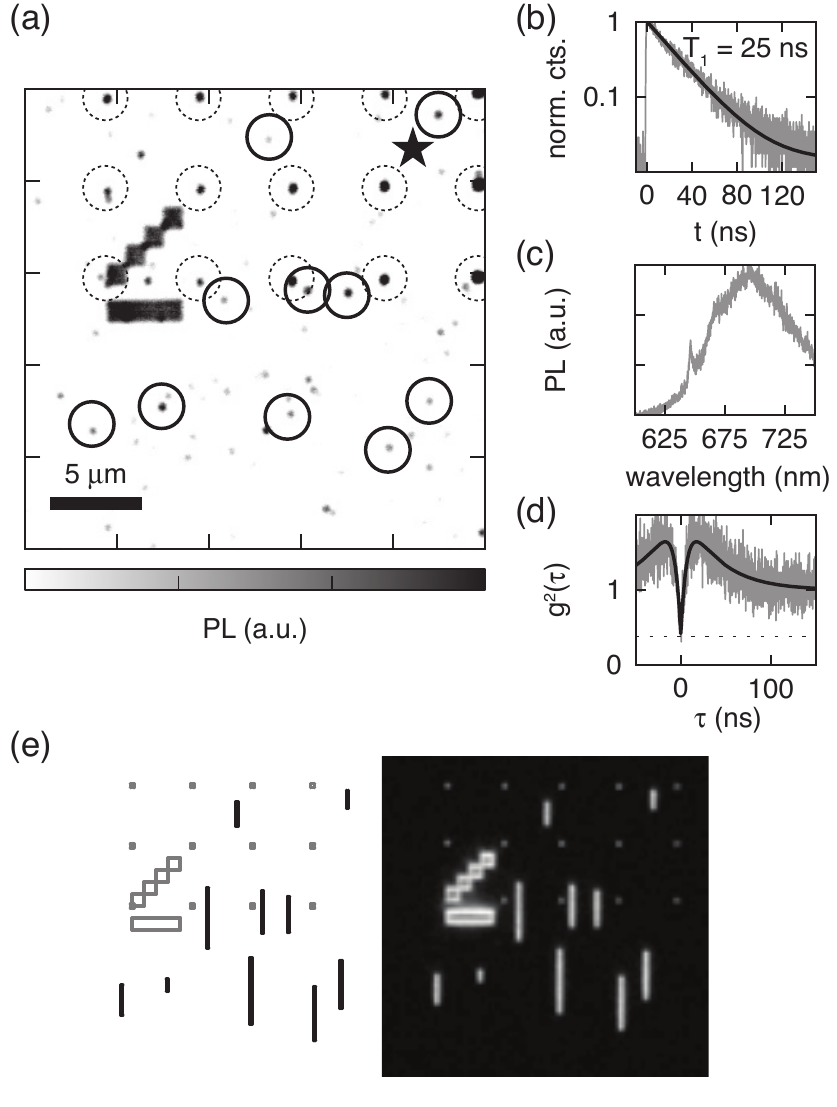}
\caption{\label{fig2}Deterministic placement of metal nanowires onto pre-selected single NV centers. {\bf a,} Confocal scan of a sample part, showing photoluminescence (PL) versus position. The spots encircled with dashed lines and the large structure on the left are gold markers. Spots encircled with solid lines are identified NV centers. {\bf b-d,} NV characterization (emitter marked with a star in panel {\bf a}). {\bf b,} Lifetime, measured with pulsed excitation. The Solid line is a fit to a single-exponential decay with offset. {\bf c,} Spectrum of the negatively charged NV centre (NV$^{-}$) with characteristic zero-phonon line at $637\,\mathrm{nm}$. {\bf d,} Antibunching in the second-order correlation function verifies that the emitter is single. Solid line is a fit to a three level model \cite{Kurtsiefer2000a}, the dashed line marks the calibrated background-level. {\bf e,} Generated pattern for electron-beam lithography (left) and optical microscopy image after fabrication (right).}
\end{figure}

We prepare our substrates by fabricating a grid of gold markers on a glass cover slip of $25\,\mathrm{mm}$ width and $0.2\,\mathrm{mm}$ thickness. We then deposit a droplet of diamond nano-crystals in aqueous solution onto the chip (mean diameter $\langle d \rangle \sim 50\,\mathrm{nm}$). In a home-built confocal microscope using green ($532\,\mathrm{nm}$) laser excitation we identify and characterize single NV centers that we find to be preferentially in the negative charge state (NV$^{-}$) (Fig.\ \ref{fig2}a-d). The glass substrate allows for optical measurements on the emitters both before and after the wire placement. We use a cover-glass corrected objective (Olympus UPlanSApo, NA=0.95) and perform all measurements from the rear side of the sample through the substrate (Fig.\ \ref{fig1}a). 

A fine grid of small gold markers allows precise location of the emitters: Fitting the PL peaks  with 2D Gaussians gives the location of both NV centers and markers within the scanning image (Fig.\ \ref{fig2}a). We then fit the position of the markers' to the real pattern on the chip. Applying the transformation obtained from the fit to the measured position of the NV centers gives their absolute location on the chip with a precision of $\sim 10\,\mathrm{nm}$. Based on this analysis we create the pattern data that is required for the EBL fabrication step (Fig.\ \ref{fig2}e). We note that both characterization measurements and the data analysis can be performed automated using straight-forward software algorithms. Therefore our fabrication approach is suited for efficient mass-fabrication of devices.

We fabricate the wires as follows. After the EBL step on a PMMA double layer resist and development in MIBK/propanol-solution we deposit the metal by electron gun evaporation and finally perform lift-off. Structure positioning with our EBL system (Vistec EBPG5000Plus HR 100) is done with an accuracy of $< 30\,\mathrm{nm}$. An optical micrograph of a finished sample is shown in Fig.\ \ref{fig2}e.

All fabrication steps have been carefully calibrated such that no NV centers are destroyed in the process, while contamination of the sample with fabrication residues remains at a minimum. NV centers close to a diamond-air interface are strongly affected by the surface of the host \cite{2012DRM....21...28G,Wolters:2012vm}. It is therefore of importance that fabrication induces no changes to the surface that bring the NV center into an unstable or non-emitting state, for instance by aggressive cleaning methods that etch away parts of the host nano-diamond. We achieve this by cleaning the sample in a modest oxygen plasma (TEPLA 100, $100\,\mathrm{W}$ microwave power, $50\,\mathrm{sccm}$ oxygen-flow) after resist development and performing lift-off in a hot acetone bath for two hours while stirring constantly. 

\section{Results and discussion}

\begin{figure}
\includegraphics{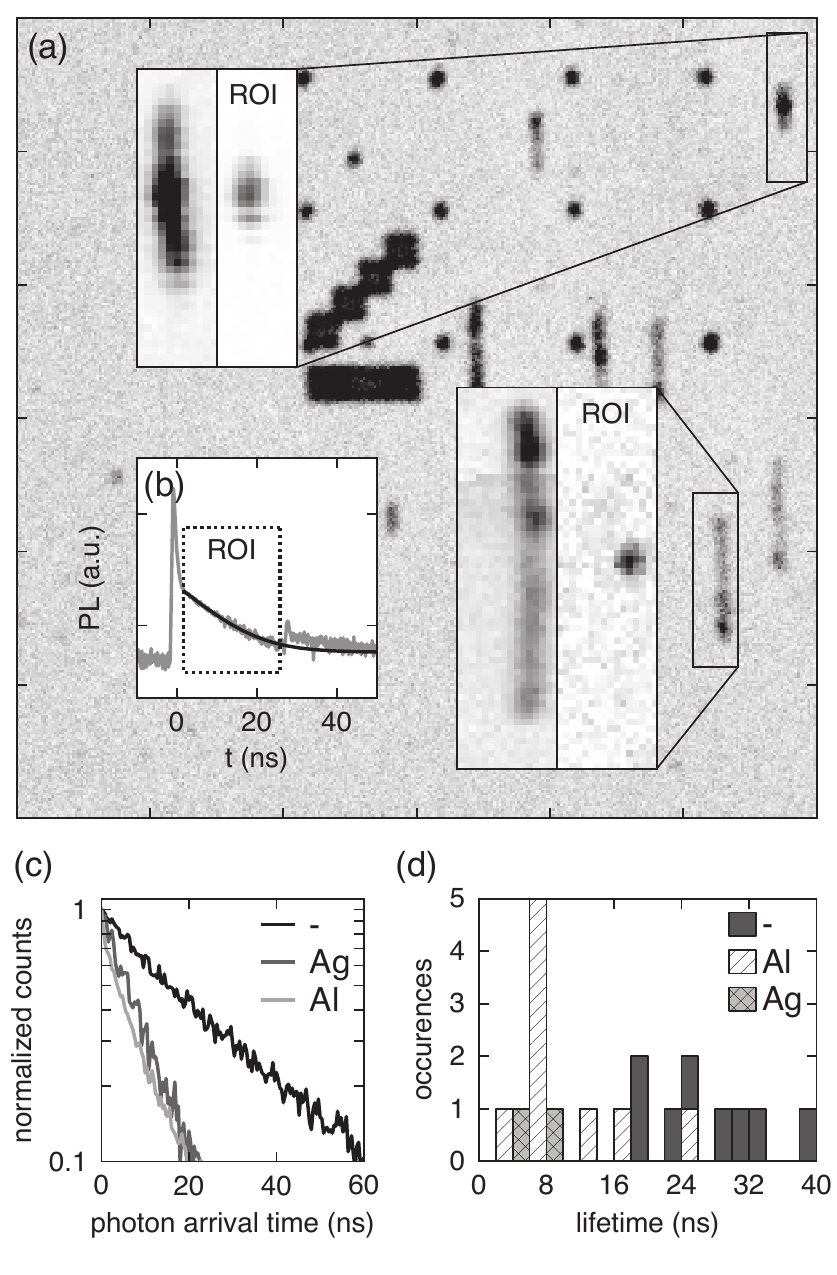}
\caption{\label{fig3}Post-fabrication characterization. {\bf a,} Confocal scan after fabrication. Insets: zoom-in onto wires placed onto single NV centers. Time-filtering the PL signal allows for localizing the NV center; {\bf b,} we filter the PL in time by defining a region of interest (ROI) such that strong, short-lived PL from metal and fabrication residues is suppressed. {\bf c,} Typical lifetimes of NV centers without ({\em ``-''}) wire and with silver ({\em ``Ag''}) and aluminum ({\em ``Al''}) wire placed on top. {\bf d,} Histogram of NV center lifetimes before and after wire fabrication.}
\end{figure}

We verify the non-destructive nature of our fabrication process by re-measuring the emitter properties. Post-fabrication confocal scans (Fig.\ \ref{fig3}a) show that photoluminescence from fabrication residues cover the signal stemming from the NV center. However, since the background PL is short-lived ($< \sim1\,\mathrm{ns}$), time-filtering the signal reveals the NV center PL, allowing for locating the emitter (Fig.\ \ref{fig3}b). With well-calibrated fabrication steps we obtain a yield of 18/19, i.e., almost all emitters can be found undamaged after processing.

Our fabrication approach is suited to observe coupling of the emitter to the structure placed on top of it. We place wires of aluminum and silver onto single NV centers and compare pre- and post-fabrication lifetimes. All investigated emitters experience a strong reduction of lifetime (Fig.\ \ref{fig3}a). Whereas the mean lifetime of the NV centers before fabrication is $T_1 = (26\pm8)\,\mathrm{ns}$, we find $T_1 = (8\pm3)\,\mathrm{ns}$ for aluminum wires (9 NV centers) and $T_1 = (7\pm2)\,\mathrm{ns}$ for silver wires (2 NV centers) placed on top (Fig.\ \ref{fig3}b).

Relating the post-measurement lifetimes to the pre-measurement lifetimes yields a Purcell factor (Eq.\ \ref{eq:purcell}) of $F_\mathrm P = 4\pm2$ for both silver and aluminum. These values are in reasonable agreement with our simulation results that predict $F_\mathrm P \sim 6$ and $8$ for silver and aluminum, respectively, at an emission wavelength of $700\,\mathrm{nm}$. Deviations are expected due to the idealized simulation geometry (Fig.\ \ref{fig1}c). Furthermore, the reduction of lifetime  is in good agreement with the values found in experiments that use random \cite{2009NatPh...5..470K} and pick-and-place assembly \cite{Huck2011,Schell:1335271}. This indicates that similar results as in these earlier works could be achieved by our fabrication scheme.

In conclusion, we have deterministically fabricated metal nano-structures onto pre-selected NV centers and observe emitter lifetime reductions compatible with coupling of the NV emission to SPP modes. Our fabrication method is robust and generally suitable for mass fabrication of hybrid emitter-plasmon devices such as antennas and waveguides. Unlike previous demonstrations of one-by-one or statistical assembly of devices, our approach is not limited to emitters embedded in nano-particles but can be used in a straight-forward way with emitters that are embedded in bulk material sufficiently close to the surface, for instance shallow-implanted NV centers in bulk diamond\cite{Toyli2010,2012ApPhL.101h2413O}. This offers the possibility to fabricate a large variety of devices in a reasonable time, and therefore to systematically study on-chip photonic circuits featuring single emitters and plasmonic nano-structures for applications.

\section*{Acknowledgements}

We thank R.~W. Heeres, T.~van der Sar and T.~H. Taminiau for helpful discussions and comments. We acknowledge support from the Dutch Organization for Fundamental Research on Matter (FOM) and the Netherlands Organization for Scientific Research (NWO).

% \bibliography{pfaff_spp.bib}


\end{document}